\documentclass[twocolumn, superscriptaddress,showpacs,aps,prl,reprint]{revtex4-1}
\usepackage{graphicx}
\usepackage{dcolumn}
\usepackage{bm}
\usepackage{blindtext}
\usepackage{times,mathptmx}
\usepackage{txfonts}
\usepackage{xcolor}

\usepackage{natbib}
\usepackage[hidelinks]{hyperref}
\hypersetup{
	colorlinks   = true, 
	urlcolor     = black, 
	linkcolor    = red, 
	citecolor   = blue 
}

\DeclareFontFamily{OMS}{oasy}{\skewchar\font48 }
\DeclareFontShape{OMS}{oasy}{m}{n}{%
	<-5.5> oasy5     <5.5-6.5> oasy6
	<6.5-7.5> oasy7     <7.5-8.5> oasy8
	<8.5-9.5> oasy9     <9.5->  oasy10
}{}
\DeclareFontShape{OMS}{oasy}{b}{n}{%
	<-6> oabsy5
	<6-8> oabsy7
	<8->  oabsy10
}{}
\DeclareSymbolFont{oasy}{OMS}{oasy}{m}{n}
\SetSymbolFont{oasy}{bold}{OMS}{oasy}{b}{n}

\DeclareMathSymbol{\smallleftarrow}     {\mathrel}{oasy}{"20}
\DeclareMathSymbol{\smallrightarrow}    {\mathrel}{oasy}{"21}
\DeclareMathSymbol{\smallleftrightarrow}{\mathrel}{oasy}{"24}

\newcommand{\monote}[1]{\textsf{{\color{blue} \scriptsize 
	}}\marginpar{{\textbf{Note}}}}

\begin{document}
	
	\title{Room temperature tunable coupling of single photon emitting quantum dots to localized and delocalized modes in plasmonic nanocavity array} 
	
	\author{Ravindra Kumar Yadav}
	\affiliation{Department of Physics, Indian Institute of Science, Bangalore 560012, India}
	\author{Wenxiao Liu}
	\affiliation{Texas A\&M University, College Station, Texas 77843, USA}
	\affiliation{Shaanxi Province Key Laboratory for Quantum Information and Quantum Optoelectronic Devices, Xi$^{\prime}$an Jiaotong University, Xi$^{\prime}$an, 710049, China}
	\author{Ran Li}
	\affiliation{Department of Materials Science and Engineering, Northwestern 
		University, Evanston, Illinois 60208, United States}
	\author{Teri W. Odom}
	\affiliation
	{Graduate Program in Applied Physics, Northwestern University, 
		Evanston, Illinois 60208, United States}
	\affiliation
	{Department of Materials Science and Engineering, Northwestern 
		Lemont, Illinois 60439, United States}
	\affiliation
	{Department of Chemistry, Northwestern University, Evanston, Illinois 60208, USA}
	\author{Girish S. Agarwal}
	\email{girish.agarwal@tamu.edu}
	\affiliation{Department of Physics, Indian Institute of Science, Bangalore 560012, India}
	\affiliation{Texas A\&M University, College Station, Texas 77843, USA}
	\altaffiliation{Department of Physics, Indian Institute of Science, Bangalore 560012, India}
	\email{girish.agarwal@tamu.edu}
	\author{Jaydeep K Basu}
	\email{basu@iisc.ac.in}
	\affiliation{Department of Physics, Indian Institute of Science, Bangalore 560012, India}
	
	\date{\today}
	
	\begin{abstract}
		Single photon sources (SPS), especially those based on solid state quantum emitters, are key elements in future quantum technologies. What is required is the development of broadband, high quantum efficiency, room temperature SPS which can also be tunably coupled to optical cavities which could lead to development of all-optical quantum communication platforms. In this regard deterministic coupling of SPS to plasmonic nanocavity arrays has great advantage due to long propagation length and delocalized nature of surface lattice resonances (SLRs). Guided by these considerations, we report experiments on the room temperature tunable coupling of single photon emitting colloidal quantum dots (CQDs) to localised and delocalised modes in plasmonic nanocavity arrays. Using time-resolved photo-luminescence measurement on isolated CQD, we report significant advantage of SLRs in realizing much higher Purcell effect, despite large dephasing of CQDs, with values of $\sim22$ and $\sim6$ for coupling to the lattice and localised modes, respectively. We present measurements on the antibunching of CQDs coupled to these modes with $g^{(2)}(0)$ values  in quantum domain providing evidence for an effective cooperative behavior. We present a density matrix treatment of the coupling of CQDs to plasmonic and lattice modes enabling us to model the experimental results on Purcell factors as well as on the antibunching. We also provide experimental evidence of indirect excitation of remote CQDs mediated by the lattice modes and propose a model to explain these observations. Our study demonstrates the possibility of developing nanophotonic platforms for single photon operations and communications with broadband quantum emitters and plasmonic nanocavity arrays since these arrays can generate entanglement between to spatially separated quantum emitters.
		
	\end{abstract}

	\maketitle
	Light sources with emitting sequential single photons with controllable quantum correlations constitute a critical component of future on-chip photonic integrated quantum technologies and for realization of quantum networks based on nanophotonics \cite {simon2017towards}. An ideal single photon emitter (SPE) should emit exactly one, indistinguishable, photon at a time which can also be used to generate entangled photon pairs while at the same time having the property of high brightness and scalability \cite {aharonovich2016solid}. In solid state quantum optical devices additional requirements include the abilities of SPE to be integrated with other components including photonic and  plasmonic cavities as well as with other optoelectronic devices \cite{koenderink2017single}. Several such SPE including atoms\cite{hijlkema2007single,ripka2018room}, ions\cite{keller2004continuous}, nitrogen-vacancy center in diamond\cite{babinec2010diamond,maletinsky2012robust}, defects in 2D materials like hexagonal Boron Nitride (hBN)\cite{sontheimer2017photodynamics,grosso2017tunable}, quantum dots (QDs) \cite{senellart2017high}, have been investigated as sources of single photons.  While it is possible to find SPEs at cryogenic temperatures which can attain some of these characteristics \cite{coles2016chirality} these can be significantly degraded at elevated temperatures which limits their applicability in various quantum technologies . A key aspect for the relatively inferior properties of SPEs near room temperature is strong optical decoherence due to various phenomena like phonon scattering \cite {utzat2019coherent}.  While no single SPE combines all the features of an ideal SPE, especially that necessary for integrated nanophotonic devices in quantum computing and related quantum technologies, a broadband single photon source with high brightness as well as photon purity and distinguishability is most desirable. 
	
	In this regard, self-assembled and colloidal quantum dots (CQDs) have been widely investigated as a potential SPE source \cite{michler2000quantum,claudon2010highly, utzat2019coherent} which can be easily integrated with other photonic and optoelectronic components and are also scalable. The biggest advantage with CQDs is their broadband spectral tunability which can be engineered by quantum confinement or composition, their very high brightness. However, most CQDs, especially for room temperature applications, show significant degradation in optical coherence due to phonon scattering and spin noise and are also impacted because of loss of indistinguishability due to dephasing or homogeneous broadening in emission linewidths \cite {utzat2019coherent}. Although single CQDs are bright single photon sources\cite{sapienza2015nanoscale} it is still not enough for them to be very effective in applications in quantum information-based technology. So, single photon emission efficiency of  CQDs needs to be enhanced which can be done by coupling these to photonic and plasmonic cavities to engineer high Purcell factor.
	
	Among the different types of cavities, plasmonic cavities are promising candidates to manipulate single photon efficiency of single CQD because of the ability to concentrate the optical field in nanoscale volume\cite{choy2011enhanced,hoang2016ultrafast,hugall2018plasmonic}. Although these cavities are quite lossy due to strong dissipation these can, nevertheless, produce quantum entanglement between two quantum emitters (QE) coupled to these cavities\cite{martin2011dissipation,nerkararyan2015entanglement,gonzalez2011entanglement,otten2015entanglement,hensen2018strong}. The loss of these plasmonic cavities can be suppressed by generating plasmonic nanocavity arrays that supports plasmonic surface lattice resonances (SLRs)\cite{kravets2018plasmonic,vak14plasmonic}, in the regime of strong coupling between the cavities. Plasmonic cavity arrays in the strong coupling regime can generate strongly delocalized modes as well as highly localized modes due to the individual cavities \cite{yadav2020strongly}. The lattice modes were utilized earlier to generate strong light-matter interaction, directional emission and lasing for high density of emitters\cite{torma2014strong,liu2016strong,shi2014spatial,yadav2020room,yang2015unidirectional,yang2015real,guo2019lasing}. Plasmonic cavity arrays have also been used, previously, to enhance single-photon emission by integrating with 2D materials like hBN\cite{tran2017deterministic,proscia2019coupling}. However, in general, while 2D materials have several advantages as SPEs, they do not have the high quantum efficiency or the broad spectral tunability that CQDs have. Thus efficient coupling of single photon emitting CQDs to these long range delocalized modes can lead to creation of a nanophotonic platform for on-chip quantum communication between such QEs through generation of long range photon entanglement with broad spectral tunability and scalability.
	
	Here, we report room temperature, tunable coupling of CQD SPEs to both localized surface plasmon resonances (LSP) and to delocalized lattice modes (SLRs) on the same plasmonic nanocavity array . By turning the lattice modes (SLRs) on and off, with the help of an index matching dielectric layer, we observe clear evidence of enhanced coupling of isolated single photon emitting CQDs to the lattice modes as compared to the localized modes in the same template through steady state and time-resolved photoluminescence (TRPL) measurements.
	We provide rigorous theoretical modeling for quantum treatment of lifetime modification to estimate the coupling strength in both the cases (LSP and SLRs). We have also shown, why a semi-classical model based on transmission spectra, hardly shows the effect of coupling due to large dephasing of  plasmonic cavity modes (LSP, SLRs) and CQDs.
	
	The results presented here are based on the use of silver nanoparticle (NP) arrays in the form of a plasmonic optical lattice. Isolated, single graded cadmium selenide/zinc sulphide (CdSe/ZnS) CQD are used as QEs inside the silver plasmonic nanocavity arrays (Supporting information(SI), Figure S1). An ultra dilute concentration of CQD solution in toluene is dispersed on plasmonic lattice using spin coating method (SI). CQDs are randomly distributed relative to individual silver NP  as shown in Fig 1(a) as a schematic. Individual CQDs were excited by 509 nm pulsed laser with a repetition rate of 1MHz. A high numerical aperture(NA=1.05) oil immersion 100x  objective was used to ensure that approximately one CQD exists in the excitation spot. 
	\begin{figure*}
		\centering
		\includegraphics[scale=0.20]{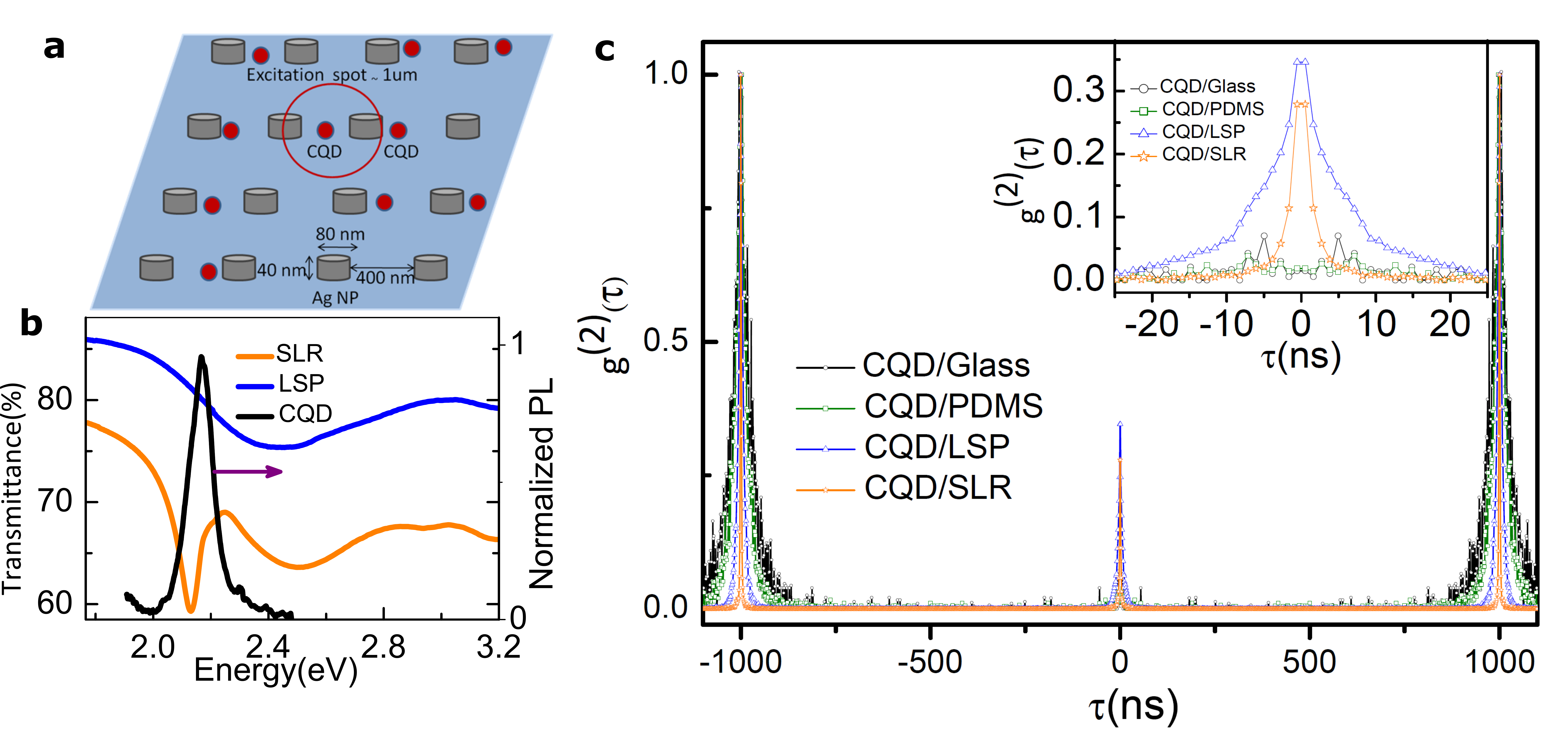}
		\caption{\label{fig1}
			(a) Schematic showing several isolated single CQDs located within the plasmonic nanocavity array.
			(b) Transmission spectra of plasmonic nanocavity array with and without homogeneous environment (PDMS),  and photoluminescence spectrum of single CQD on glass. (c) Corresponding  $g^{(2)}(\tau)$ profile for the various systems studied:CQD/Glass, CQD/PDMS, CQD/LSP, CQD/SLR respectively. (Inset shows zoomed in portion of the central peak of$g^{(2)}(\tau)$ profile around $\tau$ = 0).}     
	\end{figure*}
	
	To study the modification in life time of single CQD in the absence ( presence of LSP) and presence of lattice mode, SLRs, decay profile of  CQDs coated plasmonic lattice was collected without and with a homogeneous layer of polydimethylsiloxane (PDMS) on top of the plasmonic lattice which is known as CQD/LSP and CQD/SLR respectively. In order to study the coupling to the LSP modes in the plasmonic array we turned the collective resonances off by removing the PDMS sheet acting as a homogeneous dielectric index matching layer. Hence the CQD/LSP samples has air as the surrounding dielectric medium and does not support the collective resonances but the individual LSPs are clearly visible in the transmission spectrum. Transmission spectra for SLR and LSP  and PL spectra of single CQD  are shown in Fig 1(b). Photoluminescence spectrum of single CQD overlaps well with the transmission spectrum of SLR and partially with that of LSP mode.
	

	\begin{figure}
		\centering
		\includegraphics[scale=0.35]{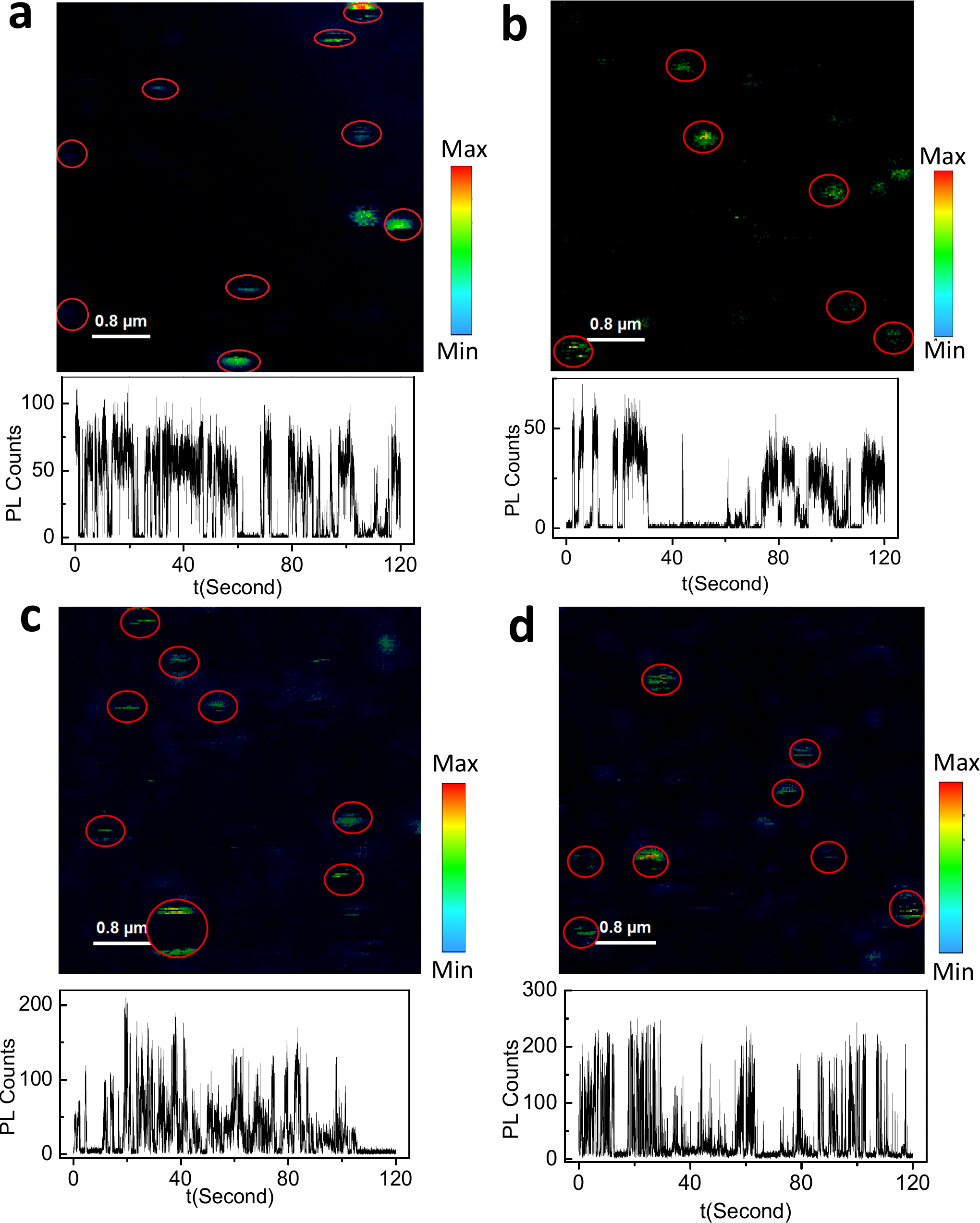}
		\caption{\label{fig2} Confocal fluorescence microscopy images (top) and corresponding PL intensity temporal (blinking) profiles (bottom) of single CQDs for  (a) CQD/Glass, (b) CQD/PDMS (c) CQD/LSP,(d) CQD/SLR, systems respectively.}     
	\end{figure}
	
	To study the single photon coupling to plasmonic nanocavity array supporting  LSPs and SLRs, we measured the blinking profile, TRPL, and second order photon correlation measurements, $g^{(2)}(\tau)$, on isolated single CQDs on glass substrate(CQD/Glass) and CQD on glass covered with PDMS sheet (CQD/PDMS) as a reference and that with individual nanocavity LSP mode (CQD/LSP) and the collective nanocavity array resonances (CQD/SLR). Here, we should note that the plasmonic nanocavity array is the same for  CQD/LSP and CQD/SLR. 
	

	Figure 1 (c)  shows corresponding second-order photon correlation measurement, $g^{(2)}(\tau)$, on single CQD which is visible as a bright spot in the PL map of each configuration (SI,Figure S3). The presence of a strong anti-bunching dip with a  value of $\sim 0.02$ which is typical of values reported for SPEs suggests $g^{(2)}(0)$ that the CQDs used in our study is a high purity single photon source. All profiles show clear anti-bunching dips at t=0 although the amplitude of the dip is larger than what was observed for bare CQD on glass. Similar increase in the amplitude of the dip has been reported in several earlier studies involving SPEs coupled to plasmonic cavities\cite{hoang2016ultrafast,tran2017deterministic,kolchin2015high} and have been ascribed to various reasons including background photons from the templates \cite {kolchin2015high}, detector dark counts \cite{tran2017deterministic} or contamination due to additional modes \cite{ihara2019superior}. Since we were able to obtain very low amplitudes with pristine CQDs on glass, detector dark counts can not be the cause of this increase. We will discuss this aspect later once we present analysis of the temporal behavior of $g^{(2)}(\tau)$ and TRPL data.  The decrease in width of $g^{(2)}(0)$ peak with amplitude,  $g^{(2)}(0)\sim 0.27$ for CQD/SLR system compared to CQD/LSP and CQD/glass or CQD/PDMS as evident in Fig. 2(d) is suggestive of an increase of coupling strength of single CQDs to the plasmonic nanocavity modes (both localized and de-localized). To extract the PL lifetime of single CQDs from $g^{(2)}(0)$, we fitted $g^{(2)}(0)$ peak with monoexponential decay function for CQD/glass and CQD/PDMS samples and with bi-exponential function for CQD/LSP, CQD/SLR samples for $\tau\ge 0$, as was evidently necessary for the respective data (SI, Figs. S4-S6). The typical values of the extracted lifetimes are shown in Table 1. Decay rate enhancements corresponding to the shorter component of extracted lifetimes, as shown in Table 1, are $\sim 6$ and  $\sim 20$ for CQD/LSP and CQD/SLRs respectively. We believe that the second longer lifetime component (weight factor is  considerably lower than the weight  factor of  shorter lifetime component) could originate from emission due to a proximal CQD which is weakly excited within the considerably enhanced field profile of the LSP and SLR modes  and hence experience sub-optimal Purcell factor enhancement. Nevertheless, this component is sufficient to degrade the spectral purity and photon distinguishability of the CQDs leading in increase in magnitude of the $g^{(2)}(\tau)$ dip for the same CQDs on the plasmonic nanocavity arrays. Temporal  filtering of the lifetime profile to eliminate the long lifetime component, as was implemented recently \cite { ihara2019superior}, can considerably restore the photon purity while still generating considerably higher decay rates and hence single photon quantum efficiency.
	\begin{table}
		\begin{tabular}{ | p{2.5cm}| p{1.1cm}| p{1.1cm}| p{1.1cm}| p{1.1cm}| p{1.1cm}|} 
			\hline
			Samples &  A$_1$ & $\tau_1(ns)$ & $A_2$ & $\tau_2 (ns)$ & $g^{(2)}(0)$  \\ 
			\hline
			CQD/Glass  & 0.02 & 29 &   &   &  0.02\\ 
			\hline
			CQD/PDMS  & 0.02 & 24  &   &   &  0.018\\ 
			\hline
			CQD/LSP & 0.22& 5 & 0.12&10.1 & 0.34 \\ 
			\hline
			CQD/SLR & 0.25 & 1.2 & 0.02 & 7.5  & 0.27\\ 	
			\hline
			
		\end{tabular}
		\caption{\textcolor{black}{Life time components extracted from  fitting $g^{(2)}(0)$ peak with $g^2(\tau)= A_1exp (-\tau/\tau_1)$ in case of CQD/Glass and CQD/PDMS and  $g^{(2)}(\tau)= A_1exp (-\tau/\tau_1)+A_2exp (-\tau/\tau_2)$ for $\tau=0$ peak, in case of  CQD/LSP, CQD/SLR as shown in Figure 1 (c).}}
		\label{table:1}
	\end{table}
	Figure 2(a)-(d) (top) shows the PL intensity map of the CQD sample on the plasmonic nanocavity array with the SLR mode turned on and off, respectively. The average distance between CQD is $\sim$1 $\mu$m which is estimated from PL map as shown in Figure 2(a)-(d).
	PL intesity maps clearly shows multiple blinking isolated single CQD in each case.
	Figure 2 (a) and (b) (bottom panels) shows blinking profiles for single CQD on glass with and without homogeneous layer respectively which shows almost similar intensity. Figure 2 (c) and (d) (bottom) shows blinking profiles of single CQDs on plasmonic array without (CQD/LSP) and with (CQD/SLR) homogeneous dielectric (PDMS) layer. The maximum intensity of CQDs in the case of CQD/LSP system is higher than CQD/glass  system while the intensity further enhances in the case of the CQD/SLR system relative to both CQD/LSP  and CQD/PDMS systems.
	This is suggestive of stronger coupling of single CQDs due to the emergence of the collective resonances in the plasmonic nanocavity arrays which is over and above any coupling to individual nanocavities in proximity to the CQD. Fluorescent enhancement factors are $\sim$ 1.5 and $\sim 4.2$ for CQD/LSP and CQD/SLR relative to respective reference samples. Another interesting observation is a drastic change in the blinking pattern. The CQD ON times becomes very short with high intensity like "photon burst" while maintaining it's quantum nature, $g^2(0)\sim 0.27 \le 0.5$ for CQD/SLR system. Blinking histograms, as shown in Fig. 3 (a)-(d),  suggest blinking is suppressed drastically with higher ON state intensity in case of CQD/SLR system which is another indication of an increased in coupling strength. Interestingly, we also observe that while the histograms for CQDs on glass show clear bimodal distribution typical for single photon emitters having an ON and OFF state\cite{cui2019colloidal}, those on the lattice display a broader distribution. This is also suggestive of additional emitting modes contributing to the emission profile due to possible contributions from remotely located and indirectly excited CQDs due to the lattice mode excitations. In the context of a recent study \cite{cui2019colloidal} involving chemically induced dimer formation of CQDs this effect was quite clearly evident both in terms of the blinking histogram as well as in the emergence of a second lifetime component in TRPL.  
	\begin{figure}
		\centering
		\includegraphics[scale=0.30]{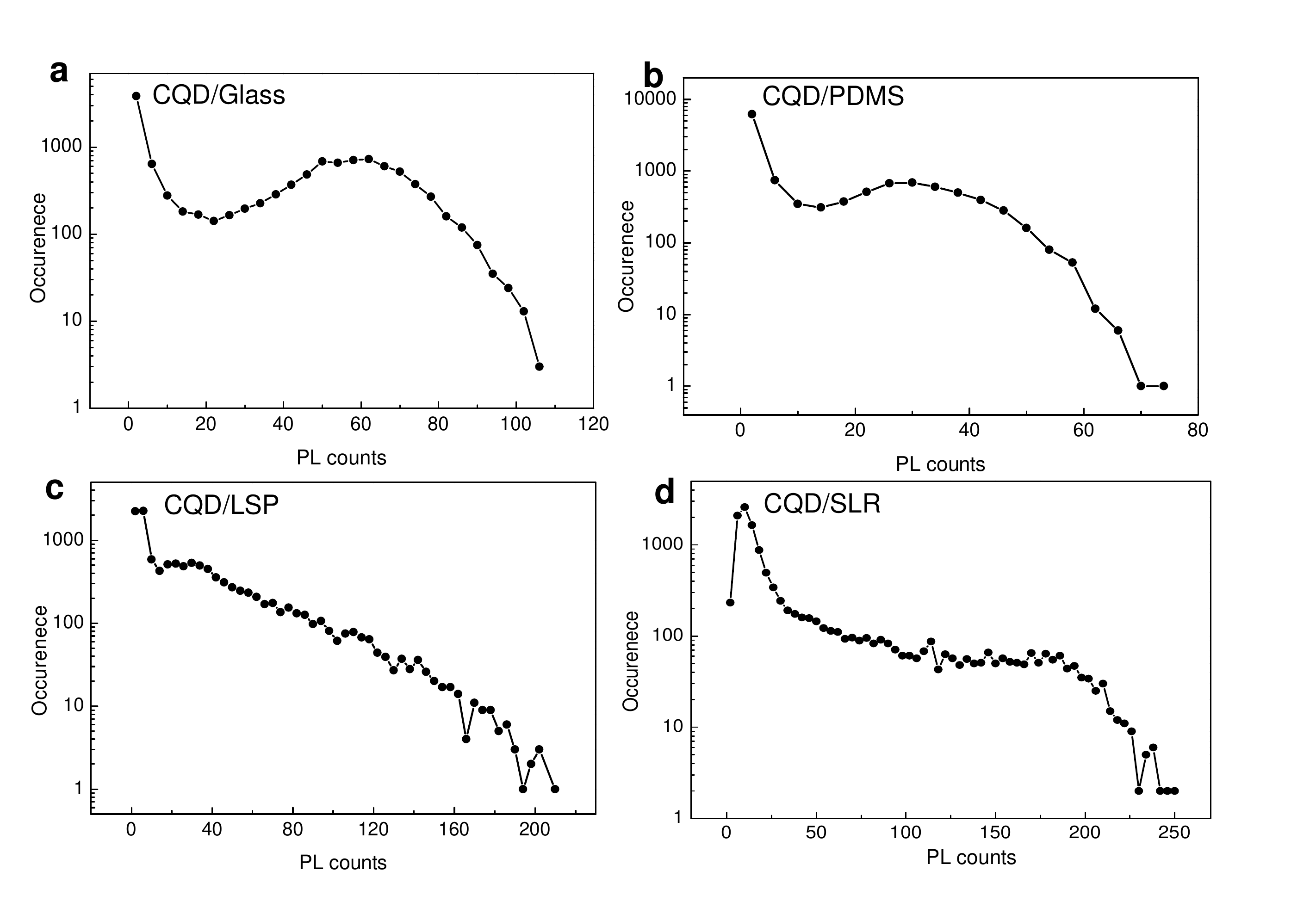}
		\caption{\label{fig3}   Typical histograms of  corresponding to blinking profiles presented in Fig 2 of single CQDs in (a) CQD/Glass, (b) CQD/PDMS (c) CQD/LSP,(d) CQD/SLR systems, respectively.}     
	\end{figure}

	\begin{figure}
		\centering
		\includegraphics[scale=.30]{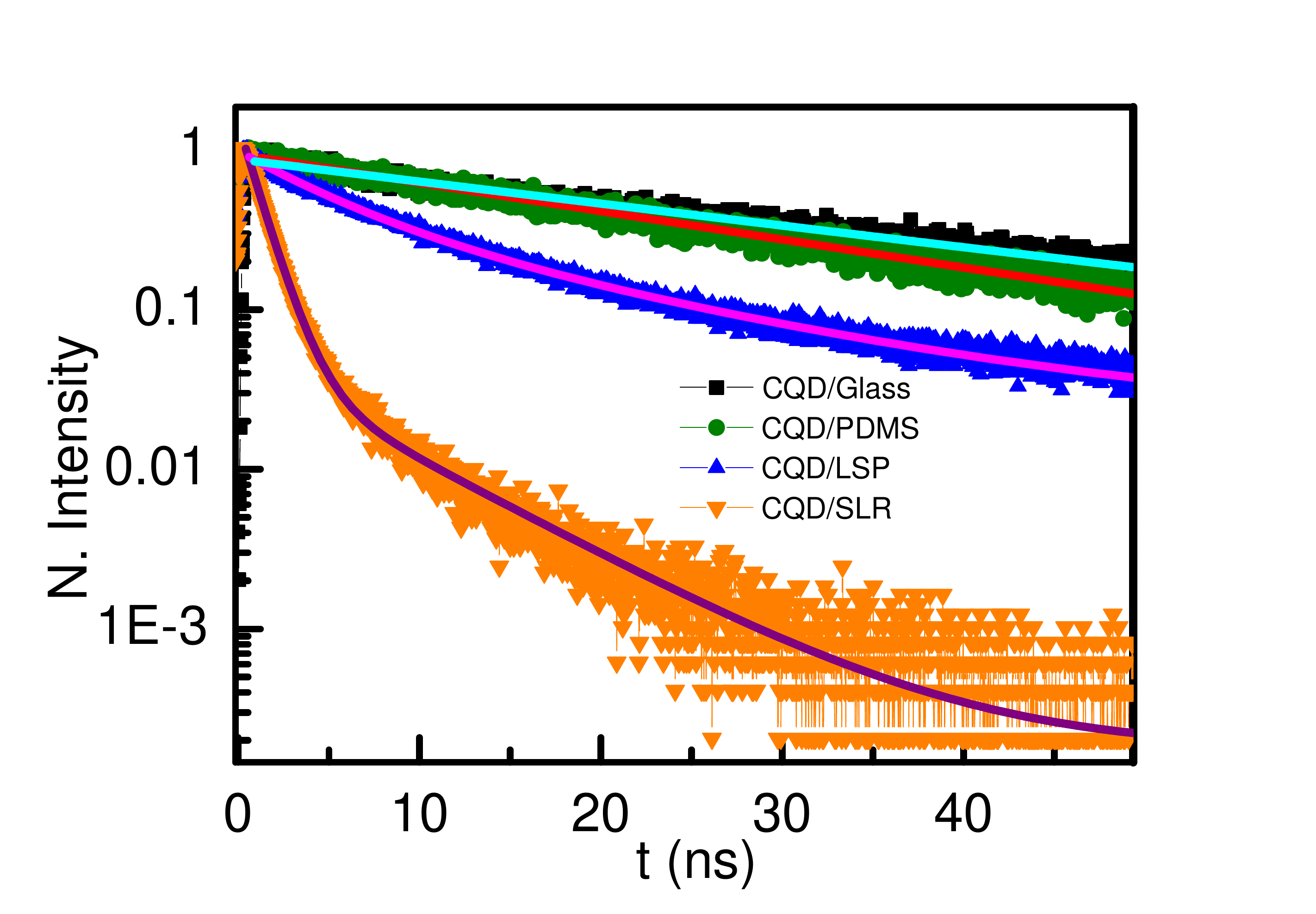}
		\caption{\label{fig4} Time resolved PL (TRPL) decay profiles of single CQDs in CQD/glass,CQD/PDMS, CQD/LSP, CQD/SLR systems.
		}     
	\end{figure}
	Further, we performed the  TRPL measurement on blinking CQD in each configuration of samples-CQD/Glass, CQD/PDMS, CQD/LSP and CQD/SLR. Decay profiles of single CQDs on glass was fitted with mono-exponential function which is similar to that reported earlier \cite{fisher2004emission}. The lifetime of CQDs for the samples-CQD/Glass and  CQD/PDMS were found to be $\sim$ 30 ns and $\sim$ 24 ns, respectively, as extracted from the mono-exponential fits. We observe a strong modification in the decay profile and lifetime of single CQD in CQD/SLR configuration relative to single CQD in CQD/LSP or CQD on glass samples. The transition from mono-exponential to bi-exponential in the decay profile of single CQD was evident in CQD/LSP and CQD/SLR system. This is consistent with the lifetime profile analysis of  the $g^{(2)}(\tau)$ data presented earlier. Table 2 shows the different lifetime components extracted from mono-exponential and bi-exponential fits of TRPL data for CQD/glass, CQD/LSP and CQD/SLR configurations. Purcell enhancement factors were estimated to be $\sim 6$ and $\sim 22$ for CQD/LSP and CQD/SLR systems respectively. Extracted lifetimes and enhancement factors estimated from $g^{(2)}(\tau)$ and TRPL measurement are consistent. However, this analysis does not provide any direct estimate of the coupling strength which is impossible to detect due to strong dephasing present in the systems studied. Hence we resort to rigorous theoretical modeling for these lifetime modifications due to coupling of LSP (localized) and SLR (delocalized) modes to single CQDs using a fully quantum formalism.
	
	\begin{table}
		\begin{tabular}{ | p{3cm}| p{1.1cm}| p{1.1cm}| p{1.1cm}| p{1.1cm}|} 
			\hline
			Samples &  A$_1$ & $\tau_1(ns)$ & $A_2$ & $\tau_2 (ns)$  \\ 
			\hline
			CQD/Glass  & 1 & 30  &   &   \\ 
			\hline
			CQD/PDMS  & 1 & 24  &   &   \\
			\hline
			CQD/LSP & 0.59 & 5 & 0.41 & 16  \\ 
			\hline
			CQD/SLR & 0.97 & 1.1 & 0.03 & 7.1 \\ 
			\hline
			
		\end{tabular}
		\caption{\textcolor{black}{Life time components extracted from fitting of decay profiles ( TRPL measurement) with bi-exponential function $I(t)= A_1exp (-t/\tau_1)+A_2exp (-t/\tau_2)$ as shown in Figure 4.}}
		\label{table:2}
	\end{table}
	
	\section{Life time data modeling: full quantum treatment}
	Here we present a full quantum treatment of the modification of the lifetimes. We have shown in the SI the inadequacy of the corresponding semiclassical studies for this purpose.
	We consider the coupling of a CQD with transition frequency $\omega_{0}$ with either the localized plasmonic mode (LSP) $\hat{b}$ or with the lattice mode (SLR) $\hat{b}$ \cite{rodriguez2011coupling,rodriguez2013surface,vak14plasmonic,guo2015controlling,laux2017single}. Let $\omega_{m}$ be the frequency of the mode. Let $G$ be the coupling constant between the CQD and the mode. We will assume the experimental arrangement so that either the SLR or LSP mode is excited. The Hamiltonian for the dot-mode system can be written as 
	\begin{equation} 
	H=\hbar \omega_{0} \hat{S}^{z}+ \hbar \omega_{m} \hat{b}^\dag \hat{b}+\hbar G (\hat{S}^{+}\hat{b}+ \hat{S}^{-}\hat{b}^\dag),
	\end{equation} 
	where the two levels of the dot $|e\rangle$ and $|g\rangle$ are used to define spin $1/2$ operators:$\hat{S}^{+}=|e \rangle \langle g|$, $\hat{S}^{-}=|g \rangle \langle e|$, $\hat{S}^{z}=(|e \rangle \langle e|-|g \rangle \langle g|)/2$. Let the states of the mode $\hat{b}$ be denoted by $|n\rangle$, i.e., $\hat{b}^{\dag}\hat{b}|n\rangle=n|n\rangle$. The life time of the dot in presence of the coupling with the mode can be obtained by examining the population of the state $|e,0\rangle$. We need to account for the life time $(2\kappa)^{-1}$ of the mode, the life time $(2\gamma_{0})^{-1}$ of the dot in the absence of the coupling to the mode and the dephasing parameter $\Gamma$ for the dot. Note that the 
	dephasing for the dot is very significant. This is quite different from the case of well known Purcell effect \cite{purcell1995spontaneous} for atoms where the dephasing is insignificant. All these in coherent processes can be accounted for in the formulae of master equations. The master equation for the density matrices of the dot-mode system is given by  \cite{agarwal2012quantum}

		$\frac{\partial \rho}{\partial t}= -\frac{i}{\hbar}[H_{1},\rho]-\kappa(\hat{b}^\dag  
			 \hat{b}\rho -2 \hat{b} \rho \hat{b}^\dag+\rho\hat{b}^\dag \hat{b})-\gamma_{0}(\hat{S}^{+}\hat{S}^{-}\rho-2$
	\begin{equation}		  
	 \hat{S}^{-} \rho \hat{S}^{+}+\rho \hat{S}^{+}  \hat{S}^{-})-\Gamma(\hat{S}^{z} \hat{S}^{z}\rho-2 \hat{S}^{z} \rho \hat{S}^{z}+\rho \hat{S}^{z} \hat{S}^{z})
\end{equation}
	
	where 
	\begin{equation} 
	H_{1}= \hbar \delta \hat{b}^\dag \hat{b}+\hbar G (\hat{S}^{+}\hat{b}+ \hat{S}^{-}\hat{b}^\dag ), \quad \delta =\omega_{m}-\omega_{0}.
	\end{equation} 
	The equation (2) is written in a frame rotating with the frequency of the dot. To obtain dynamics, we introduce the states which participate in the decay process. These are $\psi_{1}=|e,0 \rangle$, $\psi_{2}=|g,1 \rangle$, and $\psi_{3}=|g,0 \rangle$. From the master Eq.~(2), we obtain 
	\begin{equation} 
	\frac {\partial \rho_{11}}{ \partial t }=-2 \gamma_{0} \rho_{11} -i G \rho_{12}+i G \rho_{21} ,
	\end{equation}
	\begin{equation} 
	\frac {\partial \rho_{21}}{ \partial t }=i G \rho_{11}+ i \delta \rho_{21}-(\kappa+\gamma_{0}+\Gamma)\rho_{21}-i G \rho_{22},
	\end{equation}
	\begin{equation} 
	\frac {\partial \rho_{22}}{ \partial t }=i G \rho_{12}-i G \rho_{21}-2 \kappa \rho_{22}.
	\end{equation}
	The time evolution of $\rho_{11}$ will give the life time of the dot. Instead of the solving those equations in full generality we adopt a procedure of adiabatic elimination since $\kappa, \Gamma \gg \gamma_{0}$. Thus the variables $\rho_{12}$, $\rho_{22}$ decay much faster. We can set $\dot{\rho}_{21}=\dot{\rho}_{22}=0$, and then solve Eqs.~(5) and (6) for ${\rho}_{12}$, ${\rho}_{21}$, and ${\rho}_{22}$ in terms of ${\rho}_{11}$. This procedure leads to 
	\begin{equation} 
	\dot{\rho}_{11}=-2\gamma_{eff}{\rho}_{11},
	\end{equation}
	with
	\begin{equation} 
	\gamma_{eff}=\gamma_{0}+\frac{G^{2}(\kappa+\gamma_{0}+\Gamma)}{(\kappa+\gamma_{0}+\Gamma)^{2}+\delta^{2}}\{1+ \frac{G^{2}(\kappa+\gamma_{0}+\Gamma)}{\kappa[(\kappa+\gamma_{0}+\Gamma)^{2}+\delta^{2}]}\}^{-1}.
	\end{equation}
	This is the key result of our theoretical model. The effective decay rate of the dot depends on the coupling $G$ to the mode, the detuning $\delta$ from the mode, line width $\kappa$ of the mode, and the dephasing rate $\Gamma$ of the dot. For no dephasing, $\kappa \gg \gamma_{0}$,  and $\delta=0$, Eq.~(8) reduces to $\gamma_{eff}/\gamma_{0} \cong 1+G^{2}/(\kappa \gamma_{0})$, which is the celebrated Purcell result\cite{purcell1995spontaneous}, i.e., $R$. For the lattice mode (\textcolor{black}{$\delta=0.04$ ~meV} and \textcolor{black}{$\kappa, \Gamma \gg \gamma_{0}$}), we have  $\gamma_{eff}/\gamma_{0} \approx 1+G^{2}/ \left[ (\Gamma+\kappa) \gamma_{0}\right]$. From the measurements, $2\gamma_{0}=(24\times 10^{-9})^{-1}~{{s}^{-1}}$, $2\gamma_{0}+2\Gamma  \approx 90~{meV}$, \textcolor{black}{$2\kappa \sim 20 {meV}\sim (3.04 \times 10^{13})~{s}^{-1}$}, and $\Gamma/\gamma_{0}  \approx 32.82 \times 10^{5}$. Our measurements on the modification of the life time due to coupling to SLR mode gives $\gamma_{eff}/\gamma_{0} \sim 22$ and \textcolor{black}{$G\approx 1.90 \times 10^{11}~{s}^{-1}(125.3~\mu eV)$}.  We have ignored the second exponential as it's weight factor is only $3\%$. Besides the data where the second exponential is effective is highly scattered.\textcolor{black}{Second lifetime, $g$, would be lower by about 2.}
	
	Next we examine the modification in life time due to the coupling to the LSP mode. Here, $2\kappa \sim 0.439~\text{eV}$, $\omega_{m}=2.43~{eV}$, $\omega_{0}=574~{nm}=2.16~{eV}$, and \textcolor{black}{$\gamma_{eff}/\gamma_{0} \sim 6$}. This yields the coupling constant \textcolor{black}{$G \approx 2.92 \times 10^{11} {s}^{-1}(192.4\mu eV)$}. We estimate this by using the Eq.~(8). Note that according to the LSP data, for long times, the data has a decay described by an exponential almost similar to that in the absence of LSP. The weight of the exponential is about 40${\%}$. The existence of second lifetime comparable with the one in the absence of LSP mode suggests that the probability that the possibility of another dot which is very weakly coupled to the LSP. On the other hand since SLR mode has a long range, the second exponential in this case differs from the value in the absence of the SLR mode.The difference in the weight factors of the two exponentials can be understood from the fact that samples for SP and SLR studies are effectively different.
	
	\section{Theoretical modelling of the $g^{2}$ measurements}
	Next we briefly discuss theoretical modelling of $g^{(2)}$  experimental data presented in Fig 1. It is known that in the weak excitation regime, the time dependence of $g^{(2)}$ is determined by the life time of the excited state of the atom or quantum emitter\cite{agarwal2012quantum}.We would now discuss what the observed data on the $g^2$ implies and how it can be reconciled to a possible theoretical model. For the sake of simplicity we consider the case of the coupling of CQD to the lattice resonance (CQD-SLR). As discussed in the context of table 1. The data is well fitted to the form 
	\begin{equation}
	g^{(2)}(\tau)= A_1exp (-\tau/\tau_1)+A_2exp(-\tau/\tau_2)
	\end{equation}
	where $\tau_1=1.2 ns$, $\tau_2=7.5 ns$,$A_1=0.25 ns$,$A_2=0.02 ns$.The $\tau_1$ and $\tau_2$ values are quite consistent with observed values the PL data(Table 2). Let us first ignore the $A_2$ term and discuss how a value of $A_1$ about 0.25 can result.Note that if the observed $g^{(2)}$ value wore from a single data, thus the experimental value of $A_1$ would be much smaller than 0.25 i.e. The value of $A_1 $ should be in the range of the value observed for glass. We thus calculate that $A_1$ for CQD-SLR has contribution from another dot in vicinity. We propose the following scenario where the radiation collected by  the microscope objective in the state 
	\begin{equation}
	|\psi>=cos\alpha|1>+sin\alpha|2> e^{i\theta}
	\end{equation}
	\begin{figure}
		\centering
		\includegraphics[scale=0.40]{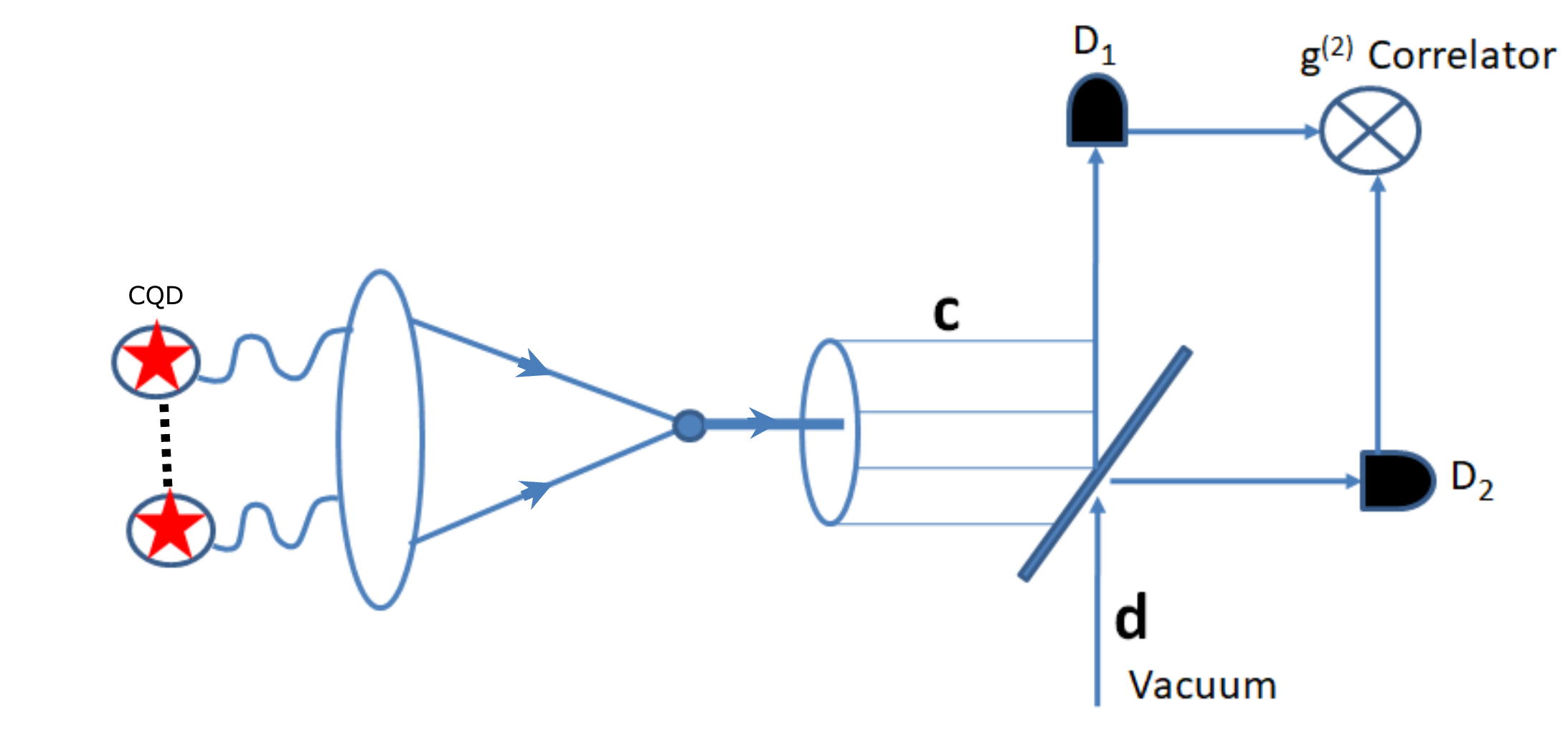}
		\caption{\label{gt} The scheme used in the calculation of $g^{(2)}$- the objective collects fluorescence from CQD's. The collected radiation is analyzed for $g^{(2)}$ as schematically shown in Fig. The value of $g^{(2)}$ depends on whether one or more dots get excited. 
		}     
	\end{figure}
	
	where we allow for the possibility that two photons are produced with probability $p=sin^2\alpha$. We include a random phase $\theta $ as the contribution from second atom is independent from the first atom. If $p=0$, then it is a true single photon situation which would have led to $g^{(2)}(0)= 0$. We now sketch a calculation of $g^{(2)}(0)$ as would be observed in the scheme shown in fig.\ref{gt}.The photonic operator at $D_1$ and $D_2$ are given by 
	\begin{equation}
	a=\frac{c+id}{\sqrt{2}}, b=\frac{d+ic}{\sqrt{2}}
	\end{equation}
	The field c is in state (9), whereas the  field d is in vacuum.Using the properties of the vacuum fields, it can be proved 
	\begin{equation}
	<a^+a>=\frac{1}{2}<c^+c>=<b^+b>, <a^+b^+ab>=\frac{1}{4}<{c^+}^2c^2>
	\end{equation}
	and hence
	\begin{equation}
	g^{(2)}(0)  =\frac{<a^+b^+ab>}{<a^+a><b^+b>}=\frac{<{c^+}^2c^2>}{<c^+c>^2} 
	\end{equation}
	The relevant expectation values in (12) can be calculated using Eq. 9 leading to 
	\begin{equation}
	g^{(2)}(0)  =\frac{2p}{(1+p)^2}=\frac{2sin^2\alpha}{(1+sin^2\alpha)^2} \end{equation}
	Thus the $g^{(2)}(0)$ would depend on the probability of second CQD contributing to the observed data. Since both CQDs are in the same neighbourhood, there coupling magnitudes are similar leading to the time scale $\tau_1$. A value of p in the range 20$\%$, would result in the experimentally observed value 0.25.Thus the scheme of fig \ref{gt} explains very well observed value.
	
	We now discuss the possible origin of the small term $A_2$ in the observed $g^{(2)}$ data  for coupling to the lattice mode.The pulse exciting  the dot, can also excite the lattice mode directly. This direct excitation would die very quickly as the lattice mode relaxation time is very short, $k\sim 10 meV$. However there is non zero probability that direct excitation of lattice mode and it's decay leads to excitation of a CQD, not with in the direct excitation region . The direct excitation(laser pulse) of the dot would be in the region where there is a significant coupling to the lattice mode, however indirect excitation would be for dots which are weakly coupled. This scenario clearly would not apply to the case of CQDs on glass and thus on glass we do not expect to see $A_2$ like contribution, which indeed is the case in experiment. In theoretical modelling this this indirect excitation of the remote dot  can be included by modifying Eq. 9 to
	\begin{equation}
	|\psi>=(cos\alpha |1> + sin\alpha .cos\beta |2>e^{i\theta_1} + sin\alpha .sin\beta|3> e^{i\theta_2 }
	\end{equation}
	where as before we included random variable $\theta_1 $ and $\theta_2 $ to account for the uncorrelated nature of the CQDs. According to the arguments presented above, the parameter $\beta$ is expected to be close to zero.The three photon contribution arises from the possibility of an indirect excitation of a CQD.The calculation of $g^{(2)}(0)$ using Eq. 14 is straight forward leading to  
	\begin{equation}
	g^{(2)}(0)=\frac{2sin^2\alpha(1+2sin^2\beta)}{[1+sin^2\alpha(1+sin^2\beta)]^2}
	\end{equation}
	The experimental values $A_2/A_1 \sim 12\sim 2sin^2\beta$ gives $sin^2\beta\sim0.04$.Thus the possibility of an indirect excitation of a CQD  is about 4$\%$ compared to the probability of the second CQD  excitation in the neighbourhood of the first CQD. In case of LSP, The $A_2$ contribution is about $50\%$ of the $A_1$ contribution. We also find that the time scale $\tau_2$ is larger.
	
	In conclusion, we provided a robust experimental scheme to simultaneous study the effect of tunable coupling of strongly dephasing, single photon emitting, CQDs to localised plasmon and delocalised lattice plasmon modes. We observed strong modification in lifetime of single CQDs, placed in plasmonic cavity array which is evident from both second order photon correlation measurements as well from the PL lifetime data. We were also able to extract the single photon coupling strengths using a fully quantum treatment of decay rate modification for both localised and delocalised lattice plasmon modes which is not possible in the transmission spectra based semi-classical models. We also report a Purcell factor of $\sim $22 for single photon coupling to lattice modes which is significantly higher than that reported earlier for single photon emitters coupled to plasmonic cavity arrays. Our study provides a nanophotonic platform using a robust scheme to integrate single CQDs with plasmonic cavity arrays, supporting long-range propagating modes, which can be used to perform efficient room temperature single photon operations with broadband quantum emitters. Such platforms involving single photon emitting quantum emitters coupled to delocalised lattice modes could find  potential applications in on-chip quantum communications and quantum information sciences.

	\section{acknowledgement}
	The authors acknowledge Indo-U.S. Science and Technol-ogy Forum (IUSSTF) for funding through a virtual centre on quantum plasmonics. The authors also acknowledge Scheme for Promotion of academic and Research Collabration (SPARC) for funding. GSA thanks the support from the Welch Foundation (No.~A-1943-20180324)and and The AFOSR award no FA9550-20-1-0366. GSA also thanks the Infosys Foundation Chair of Department of Physics, IISc Bangalore which made this collaboration successful.Authors also acknowledge some assistance in sample preparation by Weijia Wang.

\end{document}


%
%
%
%
%
%
\begin{abstract}
Here we provide  experimental methods, characterisation of samples and additional results.
\end{abstract}
\listoffigures
\section{CdSe/ZnS graded core-shell CQD Synthesis}
CdSe/ZnS graded core-shell CQD synthesized using a one-pot chemical synthesis method\cite{yadav2020room}. Chemical used for synthesis are  Cadmium oxide (CdO) (25.68 mg), zinc oxide(ZnO) (162mg), 1-octadecene(ODE) (10ml) and oleic acid (OA)  (3.52ml) and Selenium (Se) (20.5mg) and Sulfur (85mg) and trioctylphosphine(TOP) (2m). Three neck flask was loaded with Cadmium oxide (CdO) (25.68 mg), zinc oxide(ZnO) (162mg), 1-octadecene(ODE) (10ml) and oleic acid (OA)  (3.52ml) and heated to 110$^o$c and then, degassed for 20 min to create an inert environment inside the flask. After creating an inert environment inside the flask, the loaded three-neck flask was heated to 300$^o$c to get the clear yellow color solution. At 300$^o$c, the Se-S precursor was added in the three-neck flask using a syringe prepared by dissolving Se (20.5mg) and S (85mg) in 2ml TOP and temperature was kept constant for 10 min. To stop CQD's growth, the reaction mixture was cooled down to room temperature using a water bath. Synthesized CQD solution was cleaned by 1:3 ratio of chloroform and methanol using a centrifuge with speed 12000rpm for 10 minutes. Figure S1 showed the CQD solution's absorption spectra in toluene, collected using a UV-Visible spectrometer (Perkin Elmer).
\begin{figure}
	\centering
	\includegraphics[scale=0.4]{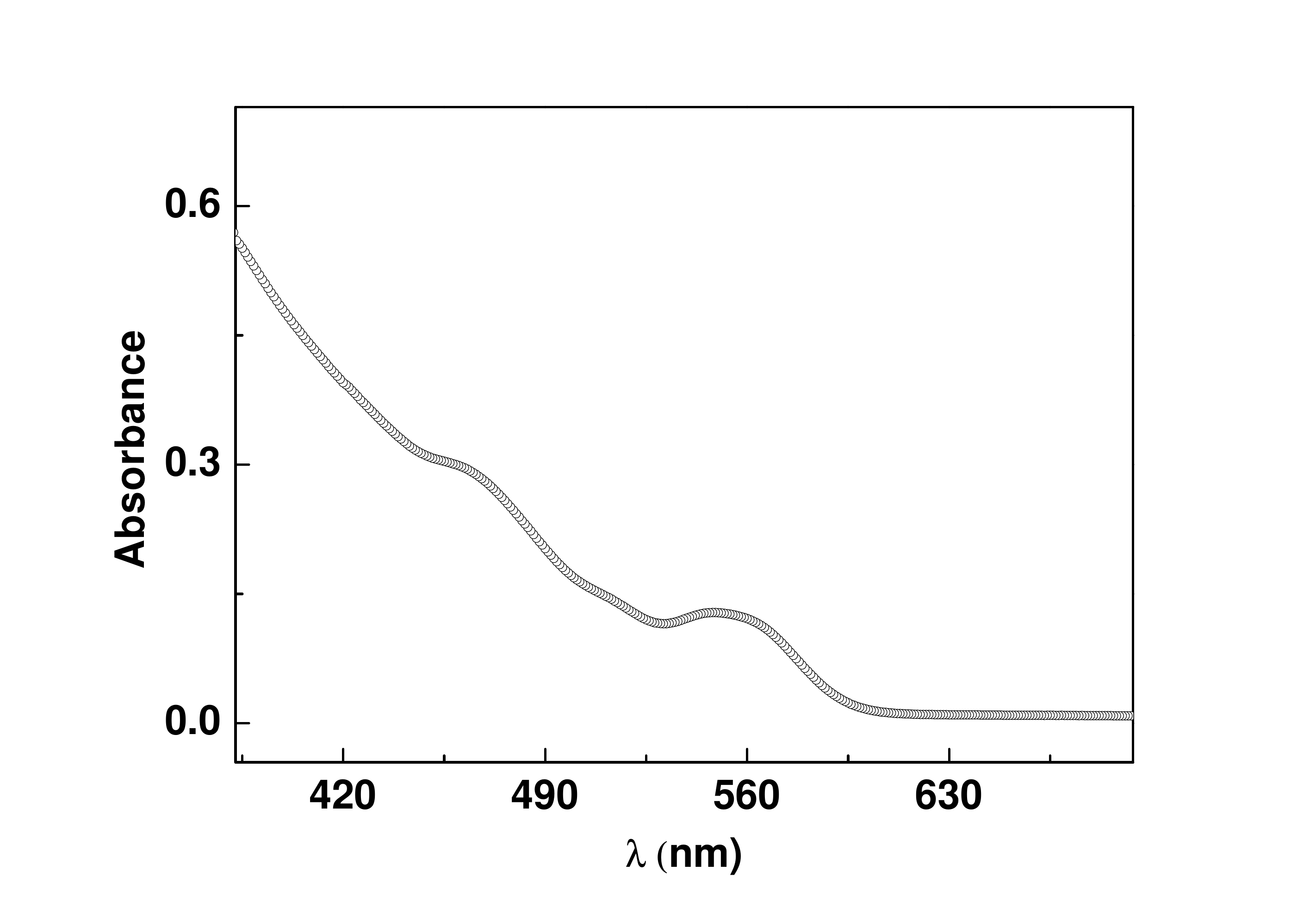}
	\caption{\label{fig1}
		Absorption spectrum of CQD solution in toluene.}     
\end{figure}
\section{Fabrication of plasmonic lattice}
Plasmonic lattice were  fabricated using  a soft nanofabrication process , known as PEEL (photolithography, etching, electron-beam deposition and lift-off)\cite{henzie2006large,yadav2020room}. Photoresist post width diameters d = 130 nm on silicon (Si) (100) wafers were produced by using a polydimethylsiloxane(PDMS) mask with square lattice spacing of 400 nm in in phase-shift photolithography. chromium (Cr) layer was deposited on  substrate using thermal vapour deposition method. Cr hole array  was produced by removing the photoresist post. Cr-Si hole array was created by Cr hole array mask on Si using reactive ion etching. To get the particle with diameter 80nm , 100 nm thick layer of gold deposited on Cr-Si hole array. Finally, Cr layer was etched  and floated Au mask on distilled water was transferred on glass substrate. Ag NP array was produced by depositing  silver on Au hole array and Au layer was removed using scotch tape.
\begin{figure}
	\centering
	\includegraphics[scale=0.6]{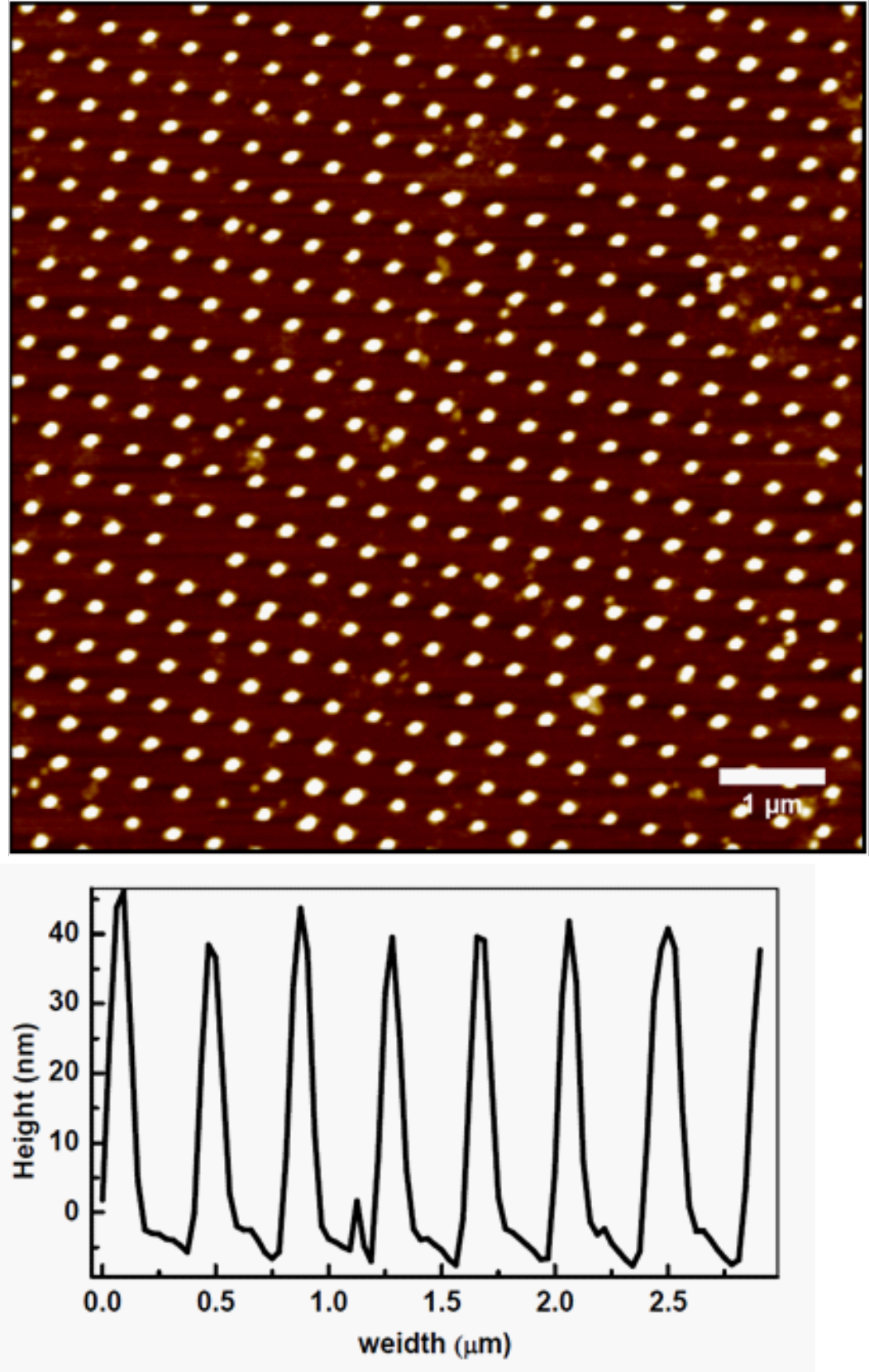}
	\caption{\label{fig1} AFM image of plasmonic lattice (top panel) and height profile on silver NP array (bottom) showing 40nm height of Ag NPs.}     
\end{figure}
\section{Deposition single isolated CQD on plasmonic lattice}
Single isolated CQDs were deposited on plasmonic lattice using a spin coating technique\cite{yadav2020room}. To isolate CQD, dilute solution of CQD in toluene $(\sim 1nM)$  was mixed with  Poly(methyl methacrylate) (PMMA) solution in toluene with a concentration of 1mg/ml. CQD/LSP sample was prepared by dispersing the CQD solution on plasmonic lattice using spin coating with speed 3000rpm for 60 second. To prepare the CQD/SLR sample, the same CQD/LSP sample was covered with a PDMS sheet to switch on SLR mode. We prepare reference sample with same CQD concentration on glass and on glass covered with PDMS sheet for CQD/LSP and CQD/SLR samples respectively. 
\section{Experimental setup for $g^{(2)}(\tau)$ measurement and time resolved photoluminescence (TRPL) measurement}$g^{(2)}(\tau)$ and TRPL measurement on single CQDs were performed on the “PicoQuant Micro Time 200 “ system. Single isolated CQD was excited with 509 nm pulse laser ( repetition rate-1MHz, power-0.1 $\mu$ W). Single CQDs were identified by collecting the PL map using Single-photon avalanche diode (SPAD). After identifying the single   CQD, $g^{(2)}(\tau)$ and blinking and TRPL measurement on single CQDs were performed for CQD/Glass, CQD/PDMS,  CQD/LSP  and CQD/SLR samples as shown in the following schematics.
\begin{figure}
	\centering
	\includegraphics[scale=0.6]{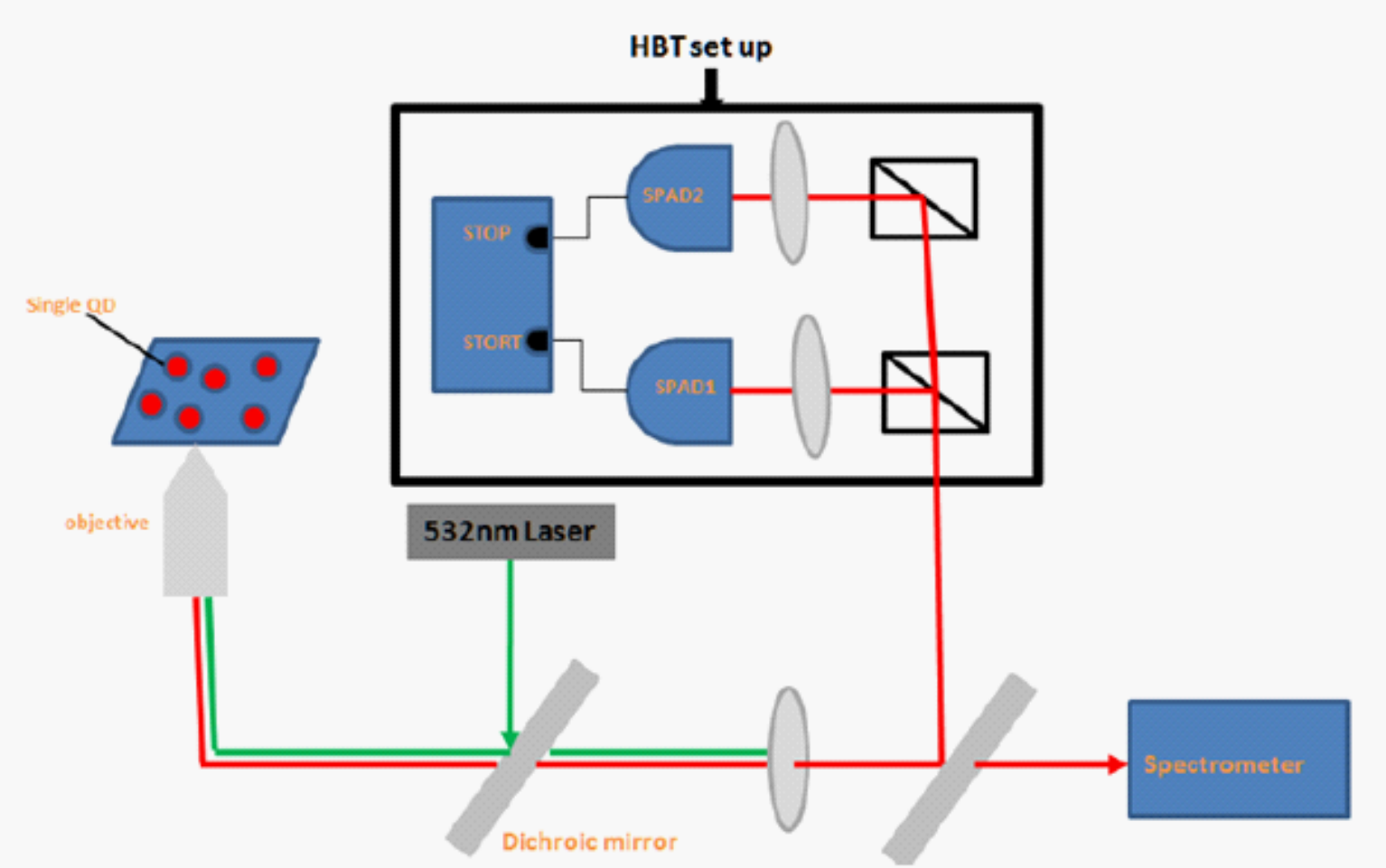}
	\caption{\label{fig1}
	Schematic of experimental set up for $g^{(2)}$  measurement}     
\end{figure}
\section{Exponential fitting of $g^{(2)}(\tau)$ and TRPL decay profiles}Here, provide the time range of fitting of various CQD configurations to extract the lifetime from  $g^{(2)}(\tau)$ and TRPL decay profile.
\section{ }
\begin{figure}
	\centering
	\includegraphics[scale=0.2]{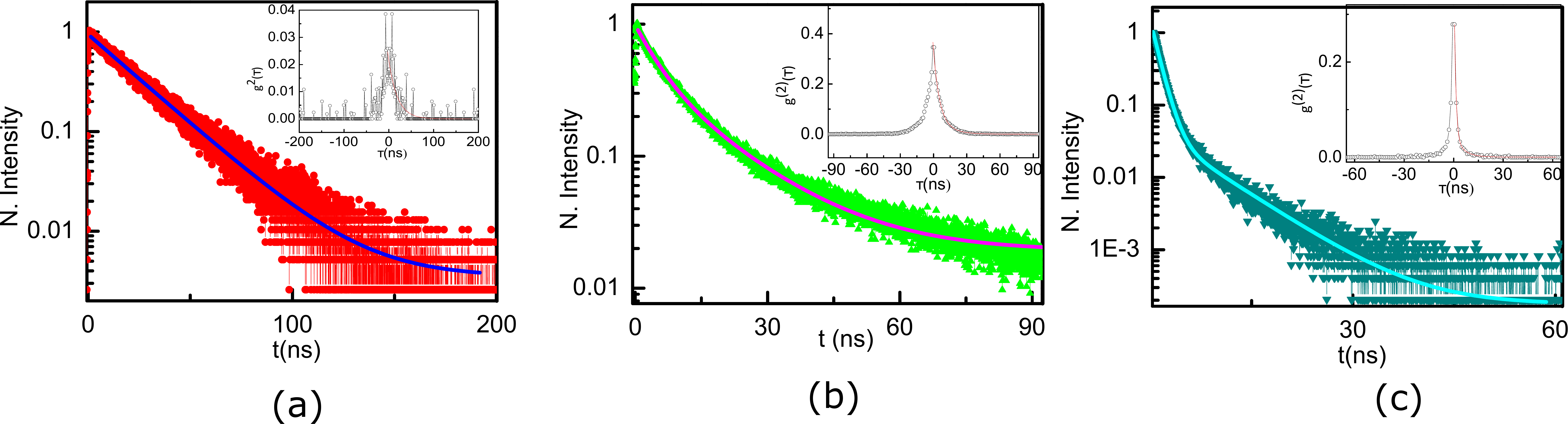}
	\caption{\label{fig1}
	Exponential fit of TRPL decay and $g^{(2)}(\tau)$ (Inset) curves for configurations (a) CQD/PDMS (b) CQD/LSP (c) CQD/SLR .}     
\end{figure}
\section{ Additional  $g^{(2)}(\tau)$ data  on CQD/plasmonic lattice }
Here, we provide the multiple g(2) data for single photon emitting CQDs  coupled to plasmonic lattice. Table 1 and 2  show the lifetime components extracted by fitting $g^{(2)}(\tau)$ to bi-exponential function for CQD/LSP and CQD/SLR  respectively.
\begin{figure}
	\centering
	\includegraphics[scale=0.6]{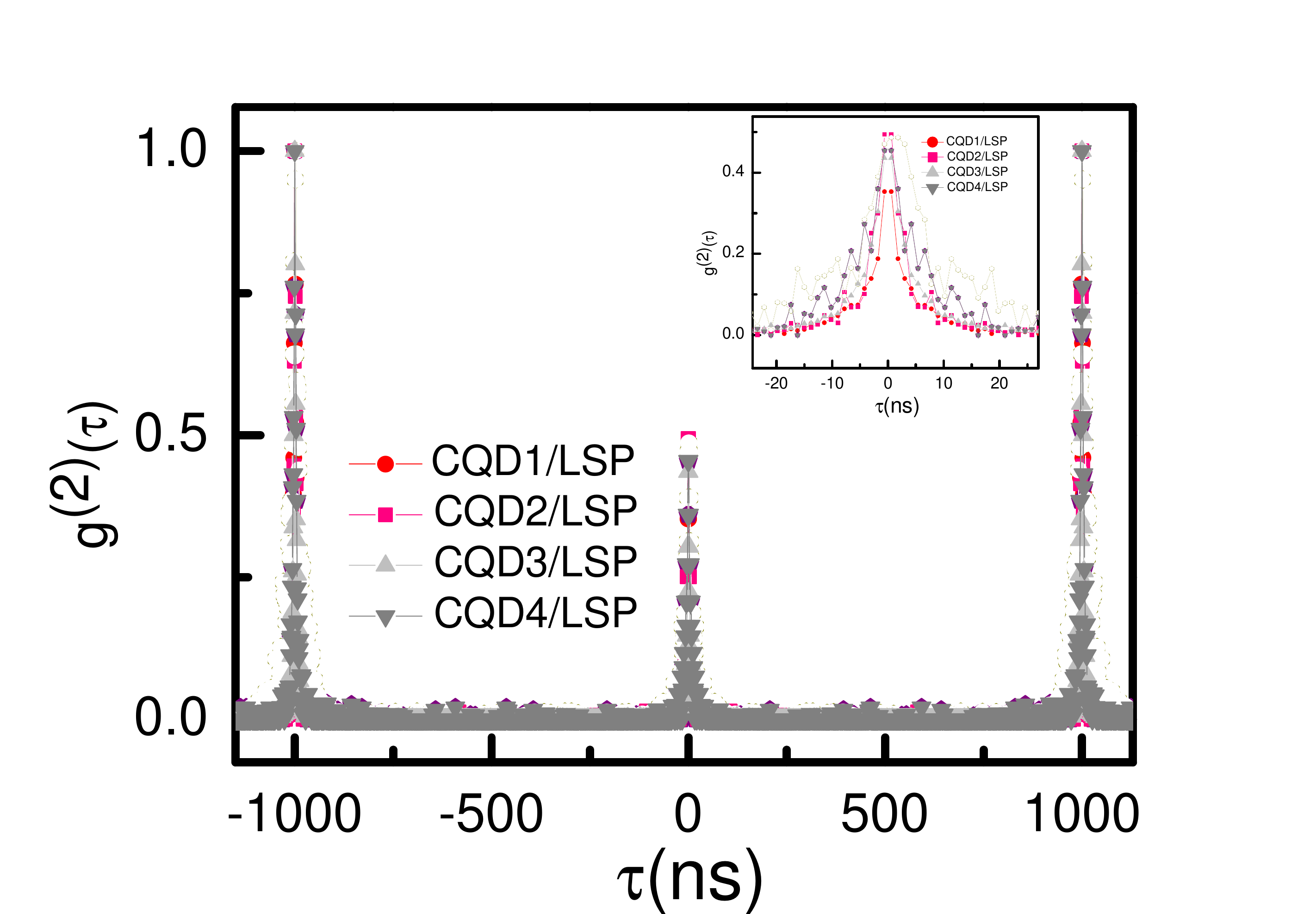}
	\caption{\label{fig1}
	$g^{(2)}(\tau)$  data for multiple CQD on CQD/LSP samples.}     
\end{figure}

\begin{table}
	\begin{tabular}{ | p{2.5cm}| p{1.1cm}| p{1.1cm}| p{1.1cm}| p{1.1cm}| p{1.1cm}| p{1.1cm}|} 
		\hline
	CQD/LSP &  A$_1$ & $\tau_1(ns)$ & $A_2$ & $\tau_2 (ns)$ & $g^{(2)}(0)$  \\ 
		\hline
		$CQD_1$  & 0.23 & 4.1 & 0.032  &  15.3 &  0.35\\ 
		\hline
			$CQD_2$  & 0.50 & 5.4  &  0.02 &  22.8 &  0.45\\ 
		\hline
		$CQD_3$ & 0.42 & 2.7 &  0.1 &  11.3 &  0.43\\ 
		\hline
		$CQD_4$ & 0.32 & 4.0  &  0.21 &  18 &  0.48\\ 
	\hline

	\end{tabular}
	\caption{Lifetime components extracted by fitting the  $g^{(2)}(\tau)$ of peak with bi-exponential function for CQD/LSP system.}
	\label{table:1}
\end{table}
\begin{figure}
	\centering
	\includegraphics[scale=0.6]{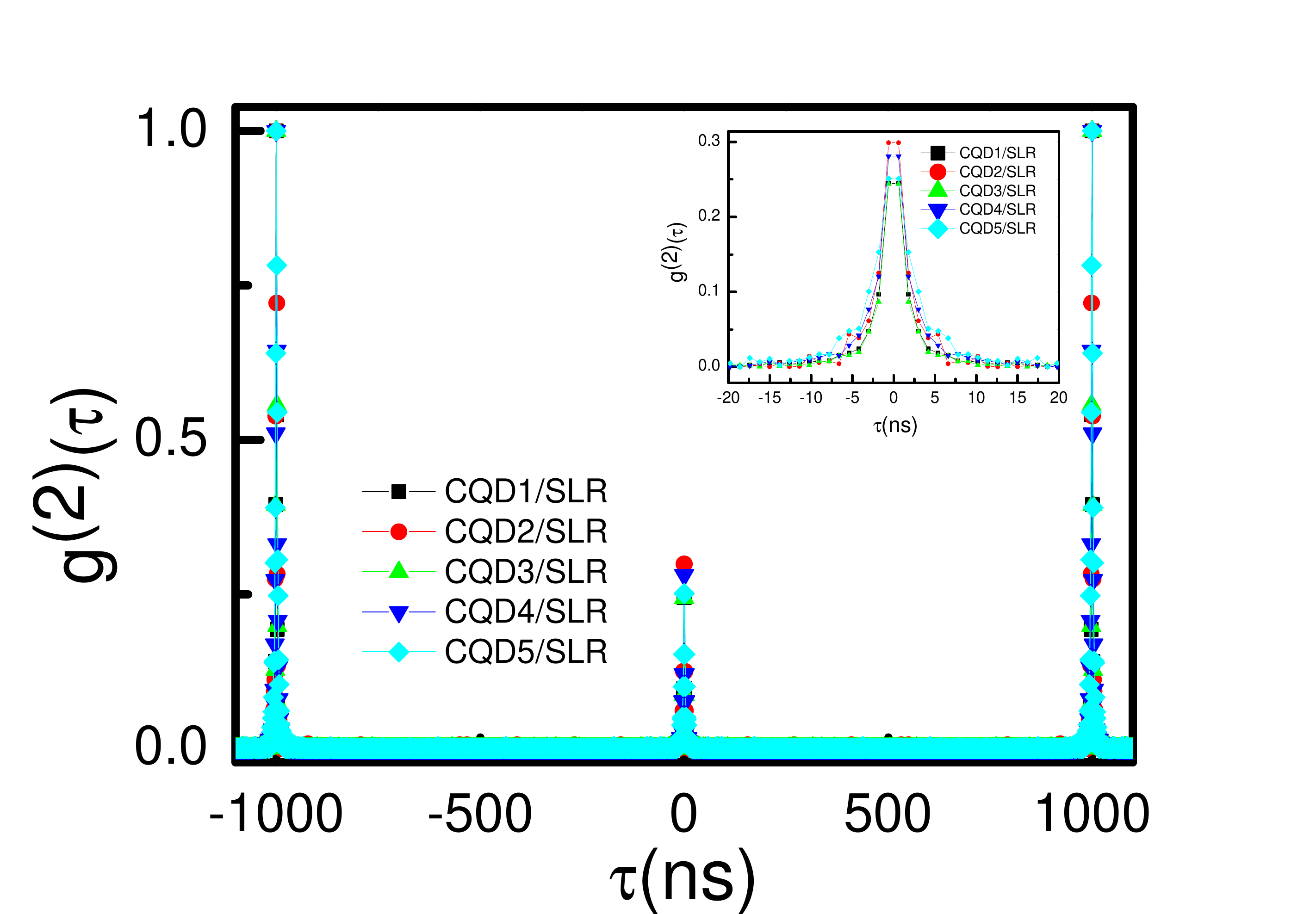}
	\caption{\label{fig1}
$g^{(2)}(\tau)$  data for multiple CQD on CQD/SLR sample.}     
\end{figure}

\begin{table}
	\begin{tabular}{ | p{2.5cm}| p{1.1cm}| p{1.1cm}| p{1.1cm}| p{1.1cm}| p{1.1cm}| }  
		\hline
	CQD/SLR &  A$_1$ & $\tau_1(ns)$ & $A_2$ & $\tau_2 (ns)$ & $g^{(2)}(0)$  \\ 
		\hline
		$CQD_1$  & 0.31 & 1.20 & 0.024  &  15.3 &  0.244\\ 
		\hline
			$CQD_2$  & 0.36 & 0.80 &  0.15 &  3.0 &  0.289\\ 
		\hline
		$CQD_3$ & 0.38 & 0.94 &  0.048 &  4.6 &  0.24\\ 
		\hline
		$CQD_4$ & 0.34& 0.97 &  0.11&  4.0 &  0.281\\ 
			\hline
		$CQD_5$ & 0.28 & 2.07  &  0.0.046 & 7.1 &  0.251\\ 
	\hline
	\end{tabular}
	\caption{Lifetime components extracted by fitting the  $g^{(2)}(\tau)$ of peak with bi-exponential function for CQD/SLR system.}
	\label{table:1}
\end{table}
\section{Semi-classical treatment: Inadequacy of transmission spectra for extracting lifetime modifications}
It is to be noticed that the factor $\gamma_\text{eff}$ is very different from what one would observe in transmission spectra. For the latter the dot-mode system would be driven by a coherent field with dots in the ground state. These spectra will be dominated by the dephasing $\Gamma$. In the following we assume the quantum dot is driven by a coherent field [$E=E_{0} \text{cos}(\omega_{l}t)$] with frequency $\omega_{l}$ and amplitude $E_{0}$.
The Hamiltonian of the dot-mode system can be given by
\begin{equation}
H=\hbar \omega_{0} \hat{S}^{z}+ \hbar \omega_{m} \hat{b}^\dag \hat{b}+\hbar G (\hat{S}^{+}\hat{b}+ \hat{S}^{-}\hat{b}^\dag)+\hbar \varepsilon(\hat{S}^{+}\text{e}^{-i\omega_{l}t}+\hat{S}^{-}\text{e}^{i\omega_{l}t}),\tag{A1}
\end{equation}
where $\varepsilon=\mu E_{0} /(2\hbar)$ is the driving strength, and $\mu$ is the dipole matrix element. The Hamiltonian in a frame rotating at $\omega_{l}$ can be written as 
\begin{equation} 
H_{2}=\hbar \Delta_{0} \hat{S}^{z}+ \hbar \Delta_{m}  \hat{b}^\dag \hat{b}+\hbar G (\hat{S}^{+}\hat{b}+ \hat{S}^{-}\hat{b}^\dag)+\hbar \varepsilon(\hat{S}^{+}+\hat{S}^{-}),\tag{A2}
\end{equation}
where 
\begin{equation}
\Delta_{0}=\omega_{0}-\omega_{l}, \quad \Delta_{m}=\omega_{m}-\omega_{l}.\tag{A3}
\end{equation} 
Using the master Equation (2) with $H_{1}$ replaced by the Eq.~(A2), we derive the mean value equations:
\begin{equation} 
\begin{split}
\frac {d \langle \hat{b} \rangle}{d t }&=-(i\Delta_{m}+\kappa)\langle \hat{b} \rangle -iG\langle \hat{S}^{-} \rangle, \\
\frac {d \langle \hat{S}^{-} \rangle}{d t }&=-(i\Delta_{0}+\gamma_{0}+\Gamma)\langle \hat{S}^{-} \rangle +2iG\langle \hat{S}^{z} \rangle \langle \hat{b} \rangle+2i \varepsilon \langle \hat{S}^{z}\rangle.
\end{split}\tag{A4}
\end{equation} 
For low excitation, small $\varepsilon$, we use $\langle \hat{S}^{z}\rangle=-\frac{1}{2}$, then we obtain in the steady state
\begin{equation}
\langle \hat{S}^{-} \rangle=\frac{-i\varepsilon}{(i\Delta_{0}+\gamma_{0}+\Gamma)+\frac{G^{2}}{i\Delta_{m}+\kappa}}.\tag{A5}
\end{equation} 
The transmission coefficient is proportional to $\langle \hat{S}^{-} \rangle/\varepsilon$. The spectral features of the transmission coefficient are given by
\begin{equation}
\Delta_{0}=\frac{i(\kappa+\gamma_{0}+\Gamma)-\delta \pm \sqrt{4G^{2}+[\delta-i(\kappa-\Gamma-\gamma_{0})}]^{2}}{2}.\tag{A6} 
\end{equation} 
The Purcell result emerges from Eq.~(A6) if we set $\delta=0$, $\Gamma=0$ and $\kappa \gg \gamma_{0}$, leading to the modification of the width $\gamma_{0}$  to $\gamma_{0} + G^{2}/\kappa$ \cite{purcell1995spontaneous}.
For SLR mode, \textcolor{black}{ $\{\Gamma, \kappa\}\gg \delta\gg \gamma_{0}$}, and $(\Gamma-\kappa) \gg \gamma_{0}$, we can obtain 
\begin{equation} 
\begin{split}
\Delta_{0}&=\frac{i(\kappa+\gamma_{0}+\Gamma)}{2} \pm \sqrt{G^2-(\kappa-\Gamma-\gamma_{0})^{2}/4} \\
&\approx
\frac{i}{2}\{\kappa+\Gamma \pm |\kappa-\Gamma| \mp \frac{G^{2}}{|\kappa-\Gamma|}\}; \quad \text{if} \quad G \ll |\kappa-\Gamma|,
\end{split}\tag{A7}
\end{equation}
In this case, \textcolor{black}{$G/(|\kappa-\Gamma|)=3.58\times 10^{-3}\ll 1$}, therefore the condition of the approximation in Eq.~(A7) can be satisfied. 
For LSP mode, $\delta=0.27~\text{eV}$, $\kappa > \Gamma$, and $(\kappa-\Gamma) \gg \gamma_{0}$, we can get
\begin{equation} 
\begin{split}
\Delta_{0}&=\frac{i(\kappa+\gamma_{0}+\Gamma)-\delta}{2} \pm \sqrt{G^2-[\delta-i(\kappa-\Gamma-\gamma_{0})]^{2}/4} \\
&\approx
\frac{i}{2}\{\kappa+\Gamma+i\delta \pm (\Gamma-\kappa-i\delta) \mp \frac{G^{2}}{\Gamma-\kappa-i\delta}\}; \quad \text{if} \quad G \ll |\Gamma-\kappa-i\delta|,
\end{split}\tag{A8}
\end{equation}
where 
\textcolor{black}{$G/(|\Gamma-\kappa-i\delta|)=5.97\times 10^{-4}$}.
Thus the transmission spectra would hardly show the effect of coupling as these would be dominated by the very large values of $\kappa$ and $\Gamma$.
\providecommand{\latin}[1]{#1}
\makeatletter
\providecommand{\doi}
{\begingroup\let\do\@makeother\dospecials
	\catcode`\{=1 \catcode`\}=2 \doi@aux}
\providecommand{\doi@aux}[1]{\endgroup\texttt{#1}}
\makeatother
\providecommand*\mcitethebibliography{\thebibliography}
\csname @ifundefined\endcsname{endmcitethebibliography}
{\let\endmcitethebibliography\endthebibliography}{}